\documentstyle[12pt]{article}
\newcommand{\bm}[1]{\mbox{\boldmath $#1$}}
\newcommand{\be}{\begin{equation}}
\newcommand{\ee}{\end{equation}}
\newcommand{\ba}{\begin{eqnarray}}
\newcommand{\ea}{\end{eqnarray}}
\newcommand{\lb}{\label}

\newcommand{\ds}{\displaystyle}
\newcommand{\ra}{\rightarrow}
\newcommand{\ol}{\overline}
\newcommand{\bb}[1]{\bibitem{#1}}
\begin{document}
\begin{titlepage}
\setcounter{page}{1}
\title{Spinning charged BTZ black holes and self-dual particle-like solutions}
\author{G\'erard Cl\'ement\thanks{E-mail:
 GECL@CCR.JUSSIEU.FR.} \\
\small Laboratoire de Gravitation et Cosmologie Relativistes
 \\
\small Universit\'e Pierre et Marie Curie, CNRS/URA769 \\
\small Tour 22-12, Bo\^{\i}te 142 \\
\small 4, place Jussieu, 75252 Paris cedex 05, France}
\bigskip
\date{\small October 12, 1995}
\maketitle
\begin{abstract}
We generate from the static charged BTZ black hole a family of spinning
charged solutions to the Einstein-Maxwell equations in 2+1 dimensions.
These solutions go over, in a suitable limit, to self-dual spinning
charged solutions, which are horizonless and regular, with logarithmically
divergent mass and spin. To cure this divergence, we add a topological
Chern-Simons term to the gauge field action. The resulting self-dual
solution is horizonless, regular, and asymptotic to the extreme BTZ black hole.
\end{abstract}
\end{titlepage}
As shown by Ba\~nados, Teitelboim and Zanelli \cite{BTZ}, the Kerr black
hole solution of (3+1)-dimensional general relativity has its counterpart
in (2+1)-dimensional general relativity with a negative cosmological
constant, a black hole solution depending on two parameters, the mass $M$
and angular momentum $J$. In the same paper, the authors of \cite{BTZ}
also gave a static charged solution to the Einstein-Maxwell equations with
a negative cosmological constant in (2+1) dimensions, the counterpart of
the Reissner-Nordstr\"om solution (the actual solution mentioned in \cite{BTZ},
a counterpart of the Kerr-Newman solution with spin $J$, is a genuine
solution only for $J = 0$ \cite{EML} \cite{K+K}). Soon after, various
spinning charged solutions, with or without horizons, were derived and briefly
discussed in a paper \cite{EML} which went largely unnoticed. Recently,
Kamaka and Koikawa \cite{K+K} found a self-dual spinning charged solution,
which they claimed to be a black hole, and to be asymptotic to the extreme
uncharged BTZ black hole; this last claim was disproved by Chan
\cite{Chan}, who pointed out that the angular momentum of the self-dual
solution diverges at spatial infinity.

The purpose of this Letter is to clarify the picture. First we show that a
simple local coordinate transformation generates the spinning charged
black hole solutions of \cite{EML} from the static charged BTZ black
hole. Then we recover the self-dual solution \cite{EML} \cite{K+K} by
taking a suitable limit in parameter space; we show that the horizon
disappears in this limit, so that the self-dual solution is really not a
black hole; moreover, it turns out to be regular. Finally, we show how the
infrared divergent behaviour of this
solution may be regularized by taking into account an additional
Chern-Simons self-coupling of the electromagnetic field, and give the
explicit regular self-dual solution \cite{Cam} in this case.

The Einstein-Maxwell action with cosmological constant $\Lambda$ is
\be \lb{1}
S = -\frac{1}{4} \int d^3x \sqrt{|g|}\,\left[ \frac{1}{4\pi G}(R + 2\Lambda)
+  g^{\mu \nu} g^{\rho \sigma} F_{\mu \rho} F_{\nu \sigma} \right]\,,
\ee
with $F_{\mu \nu} = \partial_{\mu} A_{\nu} - \partial_{\nu} A{_\mu}$. In the
case of a negative cosmological constant $\Lambda = - l^{-2}$, this action
is extremized by the two-parameter family of uncharged ($A_{\mu} = 0$) BTZ
black hole metrics
\be \lb{2}
ds^2 = N^2\,dt^2 - r^2\,(d\phi + N^{\phi}dt)^2 - \frac{dr^2}{N^2}\,,
\ee
with the lapse and shift
\be \lb{3}
N^2 = \frac{r^2}{l^2} - M + \frac{J^2}{4r^2}\,, {\hskip 12pt} N^{\phi} =
- \frac{J}{2r^2}\,,
\ee
and $M > 0$, $J^2 \leq M^2$ for the horizon to exist; other uncharged
stationary rotationally symmetric solutions are discussed in \cite{EL}.

The static charged BTZ solution \cite{BTZ} is of the form (\ref{2}) with
\ba \lb{4}
& N^2 = {\ds \frac{r^2}{l^2}} - 8\pi G Q^2\, \ln {\ds (\frac{r}{r_0})}\,, &
N^{\phi} = 0\,, \nonumber \\
& A_{\mu}\,dx^{\mu} = Q\,\ln {\ds (\frac{r}{r_0})}\, dt\,. &
\ea
This solution depends on the two parameters $Q$ (the electric charge) and
$r_0$. Note that we have not included an explicit mass term $-M$ in the
expression (\ref{4}) of $N^2$, as the value of this parameter would
depend on the scale $r_0$ \cite{BTZ}. Just as the uncharged static BTZ
solution is a black hole only for $M > 0$, the solution (\ref{4}) is a black
hole only in the domain
\be \lb{5}
r_0^2 \leq 4\pi G Q^2 l^2 {\rm e}^{-1}
\ee
of the two-parameter space \cite{EML} \footnote{Equation (\ref{5}) is derived
under the assumption that the sign of the gravitational constant $G$,
which is not fixed in three dimensions \cite{Deser84}, is positive.}.

As pointed out in \cite{Deser85} in the case $\Lambda = 0$, the local
coordinate transformation $t \ra t - \omega \phi$ generates from a
static solution of the Einstein-Maxwell equations a new, spinning solution.
Here, we shall for convenience combine this transformation with a spatial
rotation at uniform angular velocity and a length rescaling. The combined
transformation
\ba \lb{6}
t \ra t - \omega \phi\,, & \phi \ra \phi -
{\ds \frac{\omega}{l^2}}\,t\,, & r \ra \frac{\overline{l}}{l}\,r\,
\ea
(where $|\omega| < l$, and $\ol{l}^2 = l^2 - \omega^2$), generates from the
static charged solution (\ref{4}) the spinning charged solution
(previously given in \cite{EML}, equation (23)) depending on the three
parameters $Q$, $\ol{r}_0$ and $\omega$
\be \lb{7}
ds^2 = N^2\,dt^2 - K^2\,(d\phi + N^{\phi}dt)^2 - \frac{r^2}{K^2}\,
\frac{dr^2}{N^2}
\ee
with
\ba \lb{8}
& N^2 = {\ds \frac{r^2}{K^2}}\,({\ds \frac{r^2}{l^2}} -
{\ds \frac{\ol{l}^2}{l^2}}8\pi GQ^2
\ln({\ds \frac{r}{\ol{r}_0}}))\,, & N^{\phi} = - \frac{\omega}{K^2}\,
8\pi GQ^2\ln(\frac{r}{\ol{r}_0})\,, \nonumber \\
& K^2 = r^2 + \omega^2\,8\pi GQ^2\ln({\ds \frac{r}{\ol{r}_0}})\,, &
A_{\mu}\,dx^{\mu} = Q\ln(\frac{r}{\ol{r}_0})\,(dt - \omega d\phi)\,
\ea
($\ol{r}_0 = (\ol{l}/l)\,r_0$). The length rescaling in (\ref{7}) has been
chosen so that $K^2$ approaches $r^2$ at spatial infinity, and the metric
(\ref{8}) is asymptotic to the uncharged ($Q = 0$) BTZ metric, up to
logarithms. One may formally define ``mass'' and ``angular momentum''
parameters $M(r_1)$ and $J(r_1)$ for this metric by identifying, at a
given scale $r = r_1$, the values of the functions (\ref{8}) with the
corresponding uncharged BTZ values from equation (\ref{3}); however the
parameters thus defined depend on the scale $r_1$ and diverge
logarithmically when $r_1 \ra \infty$, just as observed elsewhere for
the angular momentum of the self-dual solution \cite{Chan}.

As in the static case $\omega = 0$, the spinning charged metric
(\ref{7})(\ref{8}) is a black hole only if the condition (\ref{5}), which
may here be written
\be \lb{9}
\ol{r}_0^2 \leq 4\pi GQ^2 \ol{l}^2 {\rm e}^{-1}
\ee
is satisfied. It then has two horizons, an event horizon at $r = r_+$, and a
Cauchy horizon at $r = r_-$, with $r_+ \geq r_- > \ol{r}_0$. On the other hand,
the metric function $K^2(r)$ changes sign for a certain value $r = r_c <
\ol{r}_0$ so that (as in the case of the spinning uncharged BTZ black hole)
there are closed time-like curves inside the radius $r_c$. Both the
three-geometry and the electromagnetic field are singular at $r = 0$.

The self-dual solution of \cite{K+K} (previously given in \cite{EML},
equation (29)) is obtained from (\ref{8}) by taking the limit $|\omega|
\ra l$ ($\ol{l} \ra 0$), with the other parameters $Q$
and $\ol{r}_0$ held fixed:
\ba \lb{10}
N^2 = {\ds \frac{r^2}{K^2}}\,{\ds \frac{r^2}{l^2}}\,, & N^{\phi} =
\mp {\ds \frac{l}{2 K^2}}\,M(r)\,, & K^2 = r^2 + \frac{l^2}{2}\,M(r)\,
\ea
with
\ba \lb{11}
& M(r) = 16\pi GQ^2\ln({\ds \frac{r}{\ol{r}_0}})\,,
& A_{\mu}\,dx^{\mu} = Q\ln(\frac{r}{\ol{r}_0})\,(dt \mp l d\phi)\,,
\ea
and $\pm = {\rm sign}(\omega)$ (to recover the actual form of the
one-parameter solution given in \cite{K+K}, define a new radial coordinate
$\hat r$ by $r^2 = \hat r^2 - \hat r_0^2$, where $\hat r$ and $\hat r_0 =
(4\pi GQ^2l^2)^{1/2}$ are the $r$ and $r_0$ of \cite{K+K}, and choose our
$r_0$ to have the value ${\rm e}^{-1/2} \hat r_0$). If conversely we make
on (11) the limit $Q \ra 0$, $\ol r_0 \ra 0$ with $16\pi GQ^2\ln(\ol r_0)
= -M$ held fixed, and the coordinate transformation $r^2 \ra r^2 -
(Ml^2/2)$,  we recover the extreme BTZ black hole solution, equation
(\ref{3}) with $J = \mp Ml$.

However, contrary to the extreme black hole, which has a horizon at $r^2 =
(Ml^2)/2)$ (at infinite proper distance, but finite geodesic distance),
the self-dual charged metric (\ref{10})(\ref{11}) is horizonless. The reason
is that the horizon of the spinning charged metric (\ref{8}) does not survive
the limit $\ol l \ra 0$ (according to equation (\ref{9}), it disappears below
the value $\ol l = (4\pi GQ^2)^{1/2} \ol r_0$). Apparently there still
remains in ({\ref{10}) the (now naked) singularity at $r = 0$. However,
this turns out to be at infinite geodesic distance \cite{EML}. To see
this, we note that the geodesic equations for the geometry (\ref{10})
integrate to
\be \lb{12}
\left( \frac{dr}{d\tau} \right)^2 - (l\pi_0 \pm \pi_1)^2 \frac{M(r)}{2r^2}
+ \pi_0^2 - l^{-2}\pi_1^2 + \varepsilon\, \frac{r^2}{l^2} = 0\,,
\ee
where $\tau$ is an affine parameter, $\pi_0$ and $\pi_1$ are the
constants of the motion associated with the cyclic coordinates $t$ and
$\phi$, and $\varepsilon = +1$, $0$ or $-1$ for timelike, lightlike or
spacelike geodesics. Because the function $M(r)$ in (\ref{11}) goes to
$-\infty$ as $r$ goes to zero, the only geodesics which extend to $r = 0$
are those with $\pi_1 = \mp l\pi_0$, $\varepsilon = -1$; however $r = 0$ is
then at infinite geodesic distance. The self-dual solution
(\ref{10})(\ref{11}) is thus perfectly regular, a charge-without-charge
\cite{Wheeler} solution of the three-dimensional Einstein-Maxwell equations.
Of course, the metric function
$K(r)$ still does change sign at a finite radius $r = r_c$, inside which
there are again closed timelike curves.

Returning to the general charged solution (\ref{7})(\ref{8}), we observe
that the infrared logarithmic divergences of its gravitational mass M and
angular momentum J are, just as that of the electrostatic energy of a
charge distribution in three-dimensional Minkowski spacetime, simply due
to the fact that the Abelian gauge field in (\ref{1}) is long range. These
divergences may be cured by introducing a mass term in theory. A way to do
this is to add a topological Chern-Simons term to the gauge field action,
leading to a topologically massive gauge theory \cite{DJT}. The action for
the Einstein-Maxwell-Chern-Simons theory with cosmological constant is
\be \lb{13}
S = -\frac{1}{4} \int d^3x \left\{ \sqrt{|g|}\,\left[ \frac{1}{4\pi G}(R +
2\Lambda) +  g^{\mu \nu} g^{\rho \sigma} F_{\mu \rho} F_{\nu \sigma} \right]
- \mu\,\varepsilon^{\mu \nu \rho}F_{\mu \nu}A_{\rho} \right\}\,,
\ee
where $\varepsilon^{\mu \nu \rho}$ is the antisymmetric symbol. The field
equations following from this action are
\ba \lb{14}
& & R_{\mu\nu} - \frac{1}{2}\,R\,g_{\mu\nu} = 8\pi G\,
[-F_{\mu\rho}F_{\nu}^{\,\;\rho}
+ \frac{1}{4}\,g_{\mu\nu}\,F_{\rho\sigma}F^{\rho\sigma}]
+ \Lambda\,g_{\mu\nu}\,, \\
& & \frac{1}{\sqrt{|g|}}\,\partial_{\mu}(\sqrt{|g|}\,F^{\mu\nu}) =
\frac{\mu}{2}\,\varepsilon^{\mu\rho\nu}F_{\mu\rho}\,.
\ea
Note that the topological term does not contribute to the energy-momentum
tensor. By iterating the second equation, we see that the gauge field is
massive with the mass $\mu$.

To look for stationary rotationally symmetric solutions, we use the
dimensional reduction method of \cite{EML} \cite{Cam}. The three-dimensional
metric and gauge potential are parametrized by
\ba \lb{15}
& ds^2 = \lambda_{ab}(\rho)\,dx^a dx^b -
\zeta^{-2}(\rho)\,\sigma^{-2}(\rho)\,d\rho^2\,,
& A_{\mu}\,dx^{\mu} = \psi_a(\rho)\,dx^a\,.
\ea
In (\ref{15}), $x^0 = t$, $x^1 = \phi$, $\lambda$ is the $2 \times 2$ matrix
\be \lb{16}
\lambda \equiv \left(
\begin{array}{cc}
T+X & Y \\
Y & T-X
\end{array}
\right)
\ee
with ${\rm det}\,\Lambda = -\sigma^2 = \bm{X}^2$, where
\be \lb{17}
\bm{X}^2 \equiv T^2 - X^2 - Y^2\,,
\ee
and the function $\zeta (\rho)$ allows for
arbitrary reparametrizations of the radial coordinate $\rho$. The metric
(\ref{15}) has the Lorentzian signature if the ``vector'' $\bm{X}$ is
``spacelike'' ($\bm{X}^2 < 0$).
The action (\ref{13}) reduces with this parametrization to
$I=\int d^2 x \int d\rho\,L$, with
\be \lb{18}
L = \frac{1}{2} \, [\zeta \, (-m\dot{\bm{X}}^2 +
\dot{\ol{\psi}} \, \bm{\Sigma} \cdot \bm{X} \,
\dot{\psi}) - \mu \, \ol{\psi} \, \dot{\psi} - 4 \,
\zeta^{-1} \, m\Lambda] \,,
\ee
where $\,\dot{} = d/d\rho\,$, $m= 1/16\pi G$, the real Dirac-like matrices
are defined by
\be \lb{19}
\Sigma ^0 = \left(
\begin{array}{cc}
0 & 1 \\
-1 & 0
\end{array}
\right) \, , \,\,\,
\Sigma ^1 = \left(
\begin{array}{cc}
0 & -1 \\
-1 & 0
\end{array}
\right) \, , \,\,\,
\Sigma ^2 = \left(
\begin{array}{cc}
1 & 0 \\
0 & -1
\end{array}
\right) \, ,
\ee
and $\ol{\psi} \equiv \psi^T \, \Sigma^0$
is the real adjoint of the ``spinor'' $\psi$.

Our reduced dynamical system (\ref{18}) has five degrees of freedom,
para- metrized by the coordinates $\bm{X}$, $\psi$ and the
conjugate momenta $\bm{P} \equiv \partial L/ \partial
\dot{\bm{X}}$, $\Pi^T \equiv \partial L/ \partial
\dot{\psi}$. The parametrization (\ref{15}) is invariant under an SL(2,R)
$\approx$
SO(2,1) group of transformations. The resulting invariance of the
Lagrangian (\ref{18}) under pseudo-rotations in the $\bm X$-space leads to the
conservation of the angular momentum vector
\be \lb{20}
\bm{J} = \bm{L} + \bm{S} \, ,
\ee
sum of ``orbital'' and ``spin'' contributions given by
\be \lb{21}
\bm{L} \equiv \bm{X} \wedge \bm{P} \, , \,\,\, \bm{S} \equiv
\frac{1}{2} \, \Pi^T \, \bm{\Sigma} \, \psi \, .
\ee
This last contribution may be simplified by varying the Lagrangian (\ref{18})
with respect to $\psi$, which leads to the first integrals
\be \lb{22}
\Pi^T - \frac{\mu}{2} \, \ol{\psi} = 0 \, .
\ee
(the constant right-hand side is zero in a gauge such that $\psi(\infty) = 0$).
The spin vector then takes the Dirac form
\be \lb{23}
\bm{S} = \frac{\mu}{4} \, \ol{\psi} \,
\bm{\Sigma} \, \psi \, .
\ee
Note however that, owing to the reality of the $\psi_a$, this vector is null,
\be \lb{26}
\bm{S}^2 = 0\,.
\ee
Finally, the Lagrangian (\ref{18}) is also invariant under
reparametrizations of $\rho$ (variations of $\zeta$), leading to the
Hamiltonian constraint, here written for the choice $\zeta = 1$,
\be \lb{24}
H \equiv - \frac{\bm{P}^2}{2m} - 2\mu\,\sigma^{-2}\,\bm{S}\cdot\bm{X}
+ 2m\,\Lambda = 0\,.
\ee
This Hamiltonian generates the equations of motion (again for $\zeta = 1$)
\ba \lb{25}
& & \dot{\bm{X}} = - \frac{\bm{P}}{m} \, , \nonumber \\
& & \dot{\bm{P}} = 2\mu\,\sigma^{-2}\,[ \bm{S} +
2\,\sigma^{-2}\,(\bm{S} \cdot \bm{X})\,\bm{X} ]\,, \nonumber \\
& & \dot{\bm{S}} = -2\mu\,\sigma^{-2} \, \bm{X} \wedge
\bm{S} \,\,\,\,\,\, \Longleftrightarrow \,\,\,\,\,\,
\dot{\psi} = -\mu\,\sigma^{-2} \, \bm{\Sigma} \cdot \bm{X} \,
\psi \,
\ea
(the first two equations correspond to the Einstein equations
(\ref{14}) for $\mu = a$, $\nu = b$, while the third equation
is a first integral of the gauge field equation (15)).

It is straightforward to show that the only static solutions ($Y=0$) to the
equations (\ref{25}) are vacuum solutions ($\bm{S}=0$, implying $\psi =0$).
We have obtained analytically several spinning charged solutions to these
equations. The only one which is asymptotic to a BTZ black hole follows
from the ansatz of electro-magnetic self-duality \cite{K+K}, $E = \pm B$,
which may be enforced covariantly by assuming that the scalar $F_{\mu
\nu}F^{\mu \nu}$ vanishes. With our parametrrization, this assumption reads
\be \lb{27}
\frac{1}{4}F_{\mu \nu}F^{\mu \nu} = 2\mu \sigma^{-2} \bm{S} \cdot \bm{X} =0\,,
\ee
where we have used the last equation (\ref{25}). The ansatz (\ref{27})
linearizes the second equation (\ref{25}), and the
third equation (\ref{25}) as well because, $\bm{S}$ being orthogonal to
$\bm{X}$ and null, $\bm{S} \wedge \bm{X}$ is collinear to $\bm{S}$,
\be \lb{28}
\bm{S} \wedge \bm{X} = \pm \sigma \bm{S}\,.
\ee
It also follows from (\ref{27}) and the second equation (\ref{25}) that
$\bm{J} \cdot \bm{S} =0$, implying that $\bm{J}$ is orthogonal to the null
vector $\bm{S}$, so that also $\bm{L}^2 = 0$. Then, equation (\ref{24})
yields $\dot{\sigma}^2 = 4 l^{-2}$, which integrates to $\sigma =
2l^{-1}\rho$. The integration of the full system is now straightforward.
We give the solution\footnote{There is a sign error in eq. (27)
of \cite{Cam}.} \cite{Cam} in the parametrization (\ref{7}), related
to the parametrization (\ref{15}) by
\be \lb{29}
K^2 = X - T\,,\;\;N^{\phi} = - \frac{Y}{X - T}\,,\;\; N^2 =
\frac{\sigma^2}{X - T}\,,\;\;r^2 = 2\rho\,.
\ee
This solution can be put in the self-dual form (\ref{10}) with
\ba \lb{30}
& M(r) = M - {\ds \frac{8\pi G\,Q^2}{\mu l(\mu l \mp 1)}}
{\ds ( \frac{r}{r_0} )^{\pm 2\mu l}},
& A_\mu\,dx^\mu = \pm \frac{Q}{\mu l}( \frac{r}{r_0} )^{\pm \mu l}
(dt\mp ld\phi).
\ea

The integration constant $Q$ in (\ref{30}) has been chosen so that this
solution reduces to the self-dual solution (\ref{11}) of the
Einstein-Maxwell theory in the limit $\mu \ra 0$. However the choice
of the sign $\pm$ in (\ref{10}) is not now indifferent, as the solution
(\ref{30}) explodes for $r \ra \infty$ if $\pm = {\rm sign}(\mu)$, and is
asymptotic to the extreme BTZ black hole with finite mass $M > 0$ and spin
$J = {\rm sign}(\mu)Ml$ if $\pm = - {\rm sign}(\mu)$, which we now assume.
Apart from the long-range screening of the electromagnetic field due to
the Chern-Simons term, this self-dual solution has, just as the $\mu = 0$
solution (\ref{11}), constant negative scalar curvature $R = - 6/l^2$
(this is a simple consequence of the self-duality ansatz (\ref{27})
\cite{K+K}), it is horizonless and regular (the effective mass $M(r)$ in
the geodesic equation (\ref{12}) goes to $- \infty$ for $r \ra 0$), and
admits closed timelike curves inside a certain radius $r = r_c$.

These properties ---metric asymptotic to the BTZ metric, and full
regularity of the solution--- lead us to qualify the self-dual solution
(\ref{30}) as a particle-like solution. Let us recall that similar
self-dual particle-like solutions, with the same power law behaviour for
the function $M(r)$, have been found \cite{Part} in topologically massive
gravity \cite{DJT} ---(2+1)-dimensional gravity with a gravitational
Chern-Simons term and a negative cosmological constant. This suggests that
similar solutions exist for topologically massive electrodynamics coupled
to topologically massive gravity, and in fact such a self-dual solution
has been found \cite{Cam}\cite{AM}. Let us also mention that self-dual
gravitating Chern-Simons vortices, analytical solutions to the
Einstein-Chern-Simons-Higgs equations with an eighth-order Higgs potential
\cite{Vort} have also been found. These solutions, which are also regular
and asymptotically particle-like, will be discussed elsewhere \cite{GC}.

We have generated from the static charged BTZ black hole a family of
spinning charged solutions to the Einstein-Maxwell equations. These
solutions go over, in a limiting case, to self-dual spinning charged
solutions which we have shown to be horizonless and regular. The self-dual
charged solutions to the Einstein-Maxwell-Chern-Simons equations, which we
have also constructed, share these properties and, moreover, are
asymptotic to extreme BTZ black holes.
\end{document}